\documentclass[preprint]{aastex}
\usepackage{rotating}
\renewcommand{\arraystretch}{.6}

\shorttitle{SN1993J's Light Curve} \shortauthors{Zhang Tianmeng et
 al.}

\begin{document}

\title{Optical Photometry of SN 1993J: Years 1995 to 2003}

\author{Tianmeng Zhang\altaffilmark{1}, Xiaofeng Wang\altaffilmark{1},
Xu Zhou\altaffilmark{1}, Weidong Li\altaffilmark{2}, Jun
Ma\altaffilmark{1}, Zhaoji Jiang\altaffilmark{1}, Zongwei
Li\altaffilmark{3}} \altaffiltext{1}{National Astronomical
Observatories of China, Chinese Academy of Sciences, Beijing
100012, P.R. China} \altaffiltext{2}{Department of Astronomy,
University of California, Berkeley, CA94720-3411}
\altaffiltext{3}{Department of Astronomy, Beijing Normal
University, Beijing, 100875, P.R. China}

\email{ztm@vega.bac.pku.edu.cn; wxf@vega.bac.pku.edu.cn}

\begin{abstract}
In this paper, the late-time optical photometry of supernova (SN)
1993J in M81 from Feburary 1995 to Janurary 2003 is presented. The
observations were performed in a set of intermediate-band filters
that have the advantage of tracing the strength variations of some
spectral features. SN 1993J was found to fade very slowly at late
times, declining only by $0.05\pm0.02$ mag 100 d$^{-1}$ in most of
the filters from 2 to nearly 10 yrs after discovery. Our data
suggest that the circumstellar interaction provides most of the
energy to power the late-time optical emission of SN 1993J. This
is manifested by several flux peaks seen in the rough spectral
energy distributions constructed from the multicolor light curves.
The flux peaks near 6600 \AA, 5800 \AA{} and 4900 \AA{} may
correspond to the emission lines of H$\alpha$, Na I D + He I
$\lambda$5876 and [O III] $\lambda\lambda$4959, 5007,
respectively. The evolution of these emission lines suggest a
power-law SN density model as proposed by \citet{cf94}.
\end{abstract}

\keywords{supernova: individual (SN1993J) --- techniques:
photometric}

\clearpage

\section{Introduction}
SN 1993J was visually discovered on 1993 March 28 UT by Spanish
amateur astronomer Francisco Garcia-Diaz \citep{gar93}. It
occurred in the nearby galaxy M81 (NGC 3031, $d=3.63\pm0.31$ Mpc,
Freedman et al.~1994) and reached a maximum brightness of
$M_{V}$=10.8~mag, and is the optically brightest SN in the
northern Hemisphere since SN 1954A. SN 1993J is classified as a
``Type IIb" SN: its near-maximum spectra are similar to these of
SNe II which are characterized by strong hydrogen Balmer lines,
and its nebular spectra are similar to these of SNe Ib/c that have
weak hydrogen but strong He I lines \citep{fmh93,swa93}. Models of
SN 1993J \citep{nom93, pod93, woo94} suggest that the progenitor
has lost all but a small amount of its hydrogen layers, thus the
shock-heated effective photosphere could quickly sink through the
thin H layer into the deeper He layers during the initial
expansion and the cooling phase. The light curve of SN 1993J was
unlike either of the plateau or linear version of the typical SN
II, showing two maxima similar to SN 1987A but with a faster
evolution. Except for the initial peak, the light curve bears a
strong resemblance to SN~Ib/c.

Due to its brilliant brightness and peculiarity, SN 1993J has been
extensively observed along the whole electromagnetic spectrum with
the telescopes from the ground and in orbit. Early observations of
photometry and spectroscopy in the optical bands are reported by
many authors \citep{sch93, whe93, fmb94, lew94, ben94, ric94,
pra95}. The optical photometry during the first 360 days after
explosion are published by \citet{bar95} and \citet{ric96}. The
optical spectroscopy during SN 1993J's first 2500 days is
presented by \citet{mat00a, mat00b}.

Late-time observations of SNe II have established that some SNe II
halt luminosity decline, remain optically detectable for years and
even decade after an outburst. Continued observations of these
old, evolved SNe can make constraints on the energy mechanism of
the late-time optical emission and provide clues to the evolution
of their progenitor systems.

In this paper we present the late-time photometric observations of
SN 1993J from the years 1995 to 2003. We describe our observations
and data reduction in $\S$2, and the multi-color light curves and
spectral energy distribution (SED) in $\S$3. Discussions on the
late-time energy mechanism of SN 1993J are presented in $\S$4, and
the results are summarized in $\S$5.

\section{Observations and data reduction}

Photometric observations of SN 1993J at late times have been
obtained with the 60/90 cm f/3 Schmidt telescope located at the
Xinglong station of the National Astronomical Observatory of China
(NAOC). A Ford Aerospace 2048$\times$2048 CCD camera with a 15
micron pixel size is mounted at the Schmidt focus of the
telescope. The field of view of the CCD is $58'\times58'$ with a
plate scale of 1.67 arcsec per pixel.

This telescope has a photometric system with 15 intermediate-band
(FWHM $\approx$ 200-400 \AA) filters covering a wavelength range
from 3000 \AA{} to 10000 \AA{} \citep{fan96, yan99, zhou03}. The
transmission curves of these filters are shown as the dashed
curves in Figure 1. These filters which were designed to avoid
contamination from the strongest and most variable night sky
emission lines, were used to conduct a survey project among
astronomers in Beijing, Arizona, Taipei, and Connecticut (BATC)
\citep{fan96}. We adopt the spectrophotometric AB magnitude system
in which the flux calibration is performed by observing four F
sub-dwarfs, HD 19445, HD 84937, BD+262606, and BD+174708
\citep{ok83}. The transformation equations between the broad-band
Johnson-Cousins $UBVRI$ and our intermediate-band are the
following \citep{zhou03}:

\begin{eqnarray}
U&=&b+0.6801(a-b)-0.8982\pm0.143,\nonumber\\
B&=&d+0.2201(c-d)+0.1278\pm0.076,\nonumber\\
V&=&g+0.3292(f-h)+0.0476\pm0.027,\\
R&=&i+0.1036\pm0.055,\nonumber\\
I&=&o+0.7190(n-p)-0.2994\pm0.064.\nonumber
\end{eqnarray}

We have observed SN 1993J roughly once a year using 12 filters
($d$ to $p$) since 1995. The details of the observations are given
in Table 1.

\subsection{Photometry}

To obtain proper photometry of an SN that occurs in a complicated
background (e.g. spiral arms or H II regions), observers are
usually required to take template images of the host galaxy long
after the SN has faded and do image subtraction. This method is
appropriate when the SN has a relatively fast evolution and the
template images can be obtained in a reasonable time. However,
some SNe, as the case for SN 1993J, are long-lived and evolve
slowly during the nebular phase, and the template images cannot be
obtained without the contamination of the SN light \citep{li02}.
Therefore, we have to find an alternative method to fit the galaxy
background underneath SN 1993J.

We have used a method that assumes the spiral plane around SN
1993J in M81 is a stable system which can describe by diffusive
eqution. This means, if the boundary condition for the equation is
known, we can get the flux distribution of the spiral arm at the
position of SN 1993J, and the flux of SN 1993J can be obtained by
subtracting the flux of galaxy's spiral arms from the total flux.

Before we numerically solve this Laplace equation, we do a
Gaussian smooth to reduce the noise. We then estimate the flux
distribution of M81 in a circle with a 6-pixel radius centered on
SN 1993J by interpolating the flux from the pixels outside the
circle (see the Appendix). This process is repeated until
convergency. After subtracting the estimated galaxy flux in the
circle, only the flux of SN 1993J remains. Figure 2 demonstrates
this process.

The final magnitudes of SN 1993J are measured on the subtracted
images with standard aperture photometry. We use Pipeline II (a
program developed to measure the magnitudes of point sources in
BATC images) that is based on Stetson's DAOPHOT package
\citep{ste87}.

\subsection{Calibration}

A total of 20 photometric nights were used to calibrate 30 local
standard stars in the field of SN 1993J. For each photometric
night, the afore-mentioned four standard stars were observed in a
range of airmasses, and we derive iteratively the extinction
curves and the slight variation of the extinction coefficients
$K+\Delta (UT)$. The instrumental magnitudes ($m_{\rm{inst}}$) are
calibrated by the BATC AB magnitude ($m_{\rm{BATC}}$)
\citep{zhou01} by

\begin{equation}
m_{\rm{BATC}}=m_{\rm{inst}}+[K+\Delta (UT)]\chi + C.
\end{equation}
where $\chi$ is the airmass, and $C$ is the zero point of
magnitude.

Table 1 lists the number of calibrations in each filter, while
Table 2 lists the final calibrated BATC magnitudes and their
uncertainties of the 30 local standard stars (see Figure 3 for a
finding chart of partial local standard stars). For the purpose of
clearly showing the SN image, we intercept a picture with the
field of view of $30'\times30'$ from original one degree image,
therefore only 5 out of 30 local standard stars are plotted in
Figure 3.

These local standard stars, being used to calibrate, are listed in
Table 3. The estimated error of each point is a quadrature of the
uncertainties in bias and flat-field correction, aperture
photometry, and calibrations \citep{zhou03}. The main source of
the error comes from photon noises and uncertainties in the
background subtraction.

We have also converted the BATC magnitudes of SN 1993J into the
Johnson-Cousins system ($V$, $R$, $I$ bands), in an attempt to
connect our measurements to the existing early-time broad-band
photometry. These converted $V$, $R$, $I$ magnitudes are listed in
the last three columns in Table 3. We need to point out, however,
the transformation equations we have used (also listed at the
beginning of this section) are derived from observations of normal
stars, and may bear relatively large uncertainties when apply to
the emission-dominated SN 1993J during its nebular phase.

\section{Multicolor light curves}

\subsection{Light curves during the period 1995 to 2003}

In the following sections, we adopt 1993 March 27.5 UT as the day
of explosion for SN 1993J \citep{lew94}. Our measured photometry
of SN 1993J in the $m$, $n$, $o$, and $p$ bands generally have low
quality due to the low sensitivity of these filters, and will not
be discussed hereafter. The light curves of SN 1993J in other
bands are shown in Figure 4a ($d$, $e$, $f$, and $g$ bands) and
Figure 4b ($h$, $i$, $j$, and $k$ bands) respectively.

The slow luminosity decline at late times, as listed in Table 3,
is evident for SN 1993J. The decline rates in various bands are
generally found to be $0.05\pm0.02$ 100 d$^{-1}$ during the years
from 1995 to 2003. The light curves in the $k$ and $p$ bands have
only a few points, and reliable decline rates cannot be
determined. The slow luminosity decline of SN 1993J in all the
BATC bands suggests that there is a persistent energy source
powering SN 1993J up to the most recent observations.

The light curve of $i$ band which centered on H$\alpha$, has
overall the best quality among all the bands, and displays a
distinct two-stage evolution: a relatively faster decline rate of
$0.09\pm0.01$ mag 100\,d$^{-1}$ from 700 to 2100 days after
explosion, and a relatively slower decline rate of $0.02\pm0.02$
mag 100 d$^{-1}$ from 2100 to 3600 days after explosion. The light
curves in all the other bands are consistent with a linear decline
at all times.

Figure 5 shows the evolution of SN 1993J in the $V$, $R$, $I$
passbands in the past decade. The earlier data are collected from
\citet{ric96} and \citet{bar95}, while those after 500 day were
converted from our BATC magnitudes. All the light curves exhibit
two distinct changes, the first at about 50 day after explosion,
and the second at about 400 to 500 days after explosion. The
late-time ($>$500 day) decline rate of SN 1993J in $VRI$ is far
smaller than that between 50 and 360 days after explosion
($0.06\pm0.02$mag 100 d$^{-1}$ vs 1.75$\pm$0.03mag 100 d$^{-1}$).
The dramatic slow-down after 500 day after may signal a change of
physics for the energy source (see $\S$4.2 for more discussions).

We have also constructed $R$-band magnitudes from
spectrophotometry based on the spectra published by \citet{mat00a,
mat00b}, and the result is shown as solid circles in Figure 5. The
$R$-band data from spectrophotometry is generally consistent with
those converted from the BATC magnitudes except perhaps for day
881,  which is brighter than the neighboring points by nearly 2
times. We suspect that this might be caused by they's inaccurate
flux calibration.

\subsection{SEDs at several epochs}

The late-time SED of SN 1993J can be best studied by spectroscopy
such as these done by \citet{mat00a, mat00b}. Alternatively, An
SED could be constructed from the observed fluxes in various
passbands at the same epoch. Due to the long exposure times to
observe SN 1993J in the intermediate bands, however, it is
impractical to observe all bands in a single night, and our
definition of the same epoch refers to a reference date $\pm$ 30
days. This is reasonable, considering the very slow evolution of
SN 1993J at late times.

Figure 6 shows the SEDs of SN 1993J at 700, 1000, 1423, and 3245
days after explosion, respectively. The prominent feature in the
SEDs is the strong emission from the $i$ band, which is an order
of magnitude brighter than the neighboring $h$ and $j$ bands. The
emissions in the $e$ and $g$ bands are also apparent. There is an
apparent evolution in the shape of the SED: at 700 to 1000 days
after explosion, the SED peaks at the $i$ band, while at 3245 day
after explosion, the $e$ band has the strongest emission.

Figure 1 overplots the transmission curves of the BATC system on
the spectral sequence of SN 1993J from \citet{mat00b}, and it can
be seen that our intermediate-band filters have the advantage of
properly covering some of the flat-topped lines of SN 1993J.
Because of M81's small heliocentric velocity ($-$34 km s$^{-1}$),
the $e$ filter covers the emission lines near 4900 \AA{} ([OIII]
$\lambda\lambda$4959, 5007 and H$\beta$), the $g$ filter covers
emission lines of He I $\lambda$5876 and Na I
$\lambda\lambda$5890, 5896, while the $i$ filter covers the strong
H$\alpha$ emission. Thus each of the flux peaks seen in the SEDs
corresponds to one or more strong emission lines in the spectra.

\section{Discussion}
\subsection{The late-time emission lines}

We can gain some knowledge about the evolution of some emission
lines from our multicolor intermediate-band light curves. In
theory, measuring the emission-line intensity directly from the
photometry is difficult when there is no companion knowledge of
the evolution of the line profile. This problem is somewhat
mitigated because our photometric passbands are intermediate in
size and the transmission curves are essentially flat in each
band. Moreover, the late-time spectral sequence published by
\citet{mat00b} provides some useful information on the evolution
of the line profiles. The detected emission lines, in order of
decreasing strength, are listed below.

\subsubsection{H$\alpha$ emission}

The total line flux is calculated by summing the contributions
from each unit wavelength over a range determined by the FWHM. We
estimate the FWHM of the H$\alpha$ line profile at different
epochs from the published spectra \citet{mat00b}. the FWHM of
the H$\alpha$ after 2500 day is unknown. According to the angular
expansion revealed by The Very Long Baseline Interferometry (VLBI)
observations \citep{bar02}, the expansion velocity of SN 1993J's
radio shell does not change significantly after 1600 day, we
therefore assume that the H$\alpha$ line profile does not change
after 2500 day. The effect of changing line width on the
estimation of H$\alpha$ flux will not exceed 10\% even if we allow
a decrease of 30 \AA{} for the FWHM between 2500 and 3600 days.

Adopting a distance of 3.63 Mpc \citep{fre94}  and $E(B - V) =
0.18$ mag toward SN 1993J \citep{ric94}, we estimated the
reddening corrected H$\alpha$ luminosity from the $i$-band light
curve. Figure 7 shows the results of our calculations. Note that
the calculation does not consider the contribution by the
continuum. As a result, all measurements suffer from relatively
large uncertainties. We estimate the uncertainty could be as large
as $30\%$ if we consider the fluxes measured in the $m$ to $p$
bands as the continuum.

The current H$\alpha$ emission corresponds to a luminosity of
$4.0\times10^{37}$ erg s$^{-1}$. In particular, the evolution of
the H$\alpha$ emission from SN 1993J bears a strong resemblance to
SN 1980K \citep{fhm95}. The implication of a strong persistent
H$\alpha$ emission at late times for SN 1993J will be discussed in
\S 4.2.

\subsubsection{He I $\lambda$5876/Na I $\lambda\lambda$5890, 5896
emissions}

The $g$-band light curve traces the emission lines of He I
$\lambda$5876 and/or Na I $\lambda\lambda$5890, 5896. The
luminosity of the blended lines fades very slowly from year 1995
to 2002, decreasing by only  $\sim 1.2$ mag. The total flux
measured at the last observation (2002 Mar 3 UT) is
$1.8\times10^{-14}$ erg cm$^{-2}$ s$^{-1}$, which corresponds to a
luminosity of $4.6\times10^{37}$ erg s$^{-1}$. Detailed analysis
of the spectrum on day 976 shows that He I $\lambda$5876 and Na I
$\lambda\lambda$5890, 5896 contribute approximately equally to the
broad emission feature around 5800\AA{} \citep{mat00b}. Assuming
the two components have relatively the same intensity at all
times, the above luminosity suggests that at the age of 8.9 yr
after the SN outburst, the emission line intensity of He I
$\lambda$5876 or Na I D is about half of the H$\alpha$.

\subsubsection{[O III] $\lambda\lambda$ 4959, 5007 and H$\beta$
emissions}

The $e$-band light curve traces the evolution of [O III]
$\lambda\lambda$ 4959, 5007 and H$\beta$. \citet{mat00b} suggested
that the contribution of H$_{\beta}$ to the total flux peak near
4900\AA{} may be as high as 30-40\% before day 976 (e.g. about
40\% at day 670), and decreased significantly in the following
several years (e.g. 22\% at day 1766 and 9\% at day 2454). As a
result, the late-time $e$-band light curve measures primarily the
emission from the [O III] $\lambda\lambda$ 4959, 5007 doublets. At
an earlier stage, i.e. 2-3 yrs after explosion, the [O III] line
emission is much weaker than H$\alpha$ and the He I $\lambda$5876
+ Na I D blends. At the later stage, i.e. 8-9 yrs after explosion,
however, the [O III] emission seems to gain predominance over the
other emission lines so that it is the strongest after the age of
8.9 yr. The latest $e$-band measurement (2002 Mar 3 UT) indicates
an [O III] $\lambda\lambda$ 4959, 5007 luminosity of
$5.7\times10^{37}$ erg s$^{-1}$, which is higher than the
H$\alpha$ emission at the same epoch. This is also consistent with
the spectral evolution reported by \citet{mat00b}.

\subsection{The energy sources for the late-time optical emissions}

Few SNe have had optical observations up to an age of 10 yr after
explosion. The late-time photometry, especially through more than
one filter, provides useful information on the underlying physics
for the lingering light, such as the radioactive decay of
long-lived isotopes, interaction with the circumstellar medium
(CSM), light echoes and delayed optical input by finite
recombination time. The question is how significant each component
it is.

A slow late-time decline of H$\alpha$ luminosity is in agreement
with the predicted energy input from $^{44}$Ti decay
\citep{wph89}. However, $^{44}$Ti or other long-lived radioactive
species ($^{56}$Co, $^{57}$Co, and $^{22}$Na) are unlikely to be
important energy sources for SN 1993J's optical emission at t
$\approx$ 10 yr, as the amount of $^{44}$Ti (half-life $\simeq$ 60
yr) needed to produce the current optical emissions (i.e. $i$-band
emission) is
$3.1\times10^{-3}~(D_{\rm{M81}}$/3.63~Mpc)$^{2}M_{\odot}$, far
greater than that predicted from $13-25M_{\odot}$ models
\citep{thi96} or produced in SN 1987A \citep{wph89}. Thus the
radioactive mechanism is unable to explain the bulk of the optical
emission of SN 1993J at the age of 10 yr. This means that some
other mechanisms must dominate the ionization and excitation of
hydrogen.

If the progenitor of SNe emit materials for an extended period of
time prior to exploding as a SN, they should be surrounded by a
dusty CSM. The light from the SN explosion will scatter off this
dust and produce a light echo. Since SN 1993J exploded near a
spiral arm, it is expected to illuminate the interstellar and
circumstellar material in the form of light echoes. It is likely
that a light echo component exists in the light curves of many
Type II SNe, the question is how significant the echo component is
\citep{rs00}. \citet{sc02} revealed some light echo structures
around SN 1993J by analyzing archival Hubble Space Telescope data.
They derived an observed flux from the echoes as $\sim$
4.3$\times10^{-18}$ ergs cm$^{-2}$ s$^{-1}$ \AA$^{-1}$ in years
2001, which is only about 5.4\% of the H$\alpha$ flux derived from
our i-band observations at the same time ($7.9\times10^{-17}$ ergs
cm$^{-2}$ s$^{-1}$ \AA$^{-1}$). This suggests that the light
echoes cannot account for the slow decay of the late-time light
curves of SN 1993J.

The recombination emission might give some contribution to the
optical input of supernova at late times due to the longer
recombination time scales, as \citet{kf92} showed for SN 1987A.
However, the delayed recombination should not be responsible for
SN 1993J's late-time energy. Comparison of the H$\alpha$
luminosity from model calculations \citep{kf92} and observations
of SN 1993J shows that the former is far smaller than the latter,
e.g. $\sim$1.5$\times10^{37}$erg s$^{-1}$ vs.
$\sim$2.8$\times10^{38}$ erg s$^{-1}$ at day 700. Moreover,
non-thermal emission from a young pulsar appears an equally
unlikely late-time energy source. Pulsar photoionization nebulae
should produce narrow emission lines ($\approx$ 1000 km s$^{-1}$,
Chevalier \& Fransson { }1992), in contrast to the broad emission
features seen in SN 1993J's late-time spectra \citep{mat00b}.

Chevalier \& Fransson 1994 (hereafter CF94) studied mass loss
before the explosion of a SN and suggested that the interaction of
the SN ejecta with the circumstellar wind could provide persistent
energy to power the late-time light curves. In the models CF94
studied, cool, freely expanding SN ejecta colliding with CSM from
a preexpansion stellar wind. A forward shock propagates into the
wind, while a reverse shock moves back into the ejecta. The SN
ejecta have a fairly steep density gradient, leading to a slow
reverse shock with emission at far-UV wavelengths (possibly in
X-rays, with a different gradient). This produces emission from
highly ionized species. Absorption by a shell formed at the shock
boundary can yield low-ionized lines, although these can also
originate in the ejecta themselves. The fact that all late-time,
optically detectable SNe (e.g., SNe 1957D, 1970G, 1979C, 1980K,
1993J) exhibit strong radio emission is consistent with the idea
that the main late-time energy comes from the interaction between
the expanding SN shell and slow moving CSM. Two different models
of density profile have been considered for the structure of the
wind. One is a power law, most applicable to a relatively compact
progenitor, while the other uses the density structure of a red
supergiant (RSG) from stellar evolution models. They make specific
predictions, including the line intensity ratios and line
profiles.

Table 5 list the observed line flux ratio (relative to H$\alpha$)
and the predicted values from the CF94 model, both for the
power-law wind structure and the RSG model. The measured flux
ratios from our data extend from 1.9 to 9.6 yrs after the
explosion of SN 1993J, thus providing a comparison with several
epochs of the models. The flux ratios are corrected for the
reddening of $E(B - V) = 0.18$ mag.

In general, the power-law density structure model provides better
fits to the observed line ratios than the RSG wind model.  In
particular, the decrease in H$\alpha$ emission from 1.9 to 9.9 yrs
is well produced (the dashed lines in Figure 7). Assuming Na I D
contributes to half of the intensity of the emission at 5800 \AA,
its observed flux matches the model prediction at 2 to 5 yrs after
explosion, but falls short of the prediction at 10 yr (only about
50\%). Part of the reason for this discrepancy could be caused by
the changing ratio between Na I D and He I $\lambda$5876 at very
late times. The power-law density structure model of CF94, for
example, predicts the Na I D line would be twice strong as He I
$\lambda$5876 at late times. Allowing a relatively large
contribution of H$\beta$ to the [O III] $\lambda\lambda$4959,
5007+H$\beta$ blend before day 1766 ($20\% - 40\%$), the observed
[O III] $\lambda\lambda$4959, 5007 flux matches the model
prediction at 2 to 5 yrs after explosion. At 6.7 yr, the
contribution of H$\beta$ to [O III]+ H$\beta$ blend is less than
10\%. If we assume that all flux at 5000 \AA{} are caused by [O
III] at 10 yr, we found that the observed [O III] flux does not
deviate significantly from the power-law model predictions. The
RSG wind model (the last column in Table 5B) generally produces
much stronger lines of [O III] $\lambda\lambda$4959, 5000 and much
weaker Na I D lines than observed.

\section{Conclusions}

We present intermediate-band photometry of SN 1993J from 692 to
3620 days after discovery, greatly extending the coverage of its
evolution in the optical bands. The intermediate-band light curves
show a very slow decline after day 700, fading at $0.05\pm0.02$
mag 100 d$^{-1}$.

We constructed the SEDs of SN 1993J from the measured flux of the
multicolor light curves. The SEDs show flux peaks near 4900 \AA,
5800 \AA,  and 6600 \AA{} in order of increasing strength, which
are associated with the line emissions of [O III]
$\lambda\lambda$4959, 5007, Na I D/He I $\lambda$5876, and
H$\alpha$,  respectively. The [O III] doublet emission gains
dominance over time so that it becomes the strongest emission at
day 3245.

Several emission lines seen in the SEDs of SN 1993J and their
evolution provide evidences that the interaction of the ejecta
with the CSM is the primary energy source to power the late-time
optical emission. We also found that the power-law density profile
model of the interaction model gives quantitative agreement with
the observations. In particular, the line ratio of [O III] doublet
and Na I D relative to H$\alpha$ are well reproduced by the model.

In the years to come, it will still be interesting to monitor the
evolution of the optical emission of SN 1993J. In particular, the
strength of [O III] $\lambda\lambda$4959, 5007 will increase
continuously according to the circumstellar interaction model.
Continued observations of old SNe such as SN 1993J offer us an
opportunity to study the transition of evolved SNe into supernova
remnants.

\acknowledgments  We are grateful to Mr. Yang Yanbin for his
helpful discussions on reduction of the photometric data.
Financial support for this work has been provided by the National
Science Foundation of China (NSFC grant 10303002; 10173003) and
National Key Basic Research Science Foundation (NKBRSF
TG199075402).

\appendix
\section{Appendix: The numerical solution of the  Laplace equation}

The Laplace equation is:
\begin{equation}
\nabla^{2}U=\frac{\partial ^{2}U}{\partial x^{2}}+\frac{\partial
^{2}U}{\partial y^{2}}=0.
\end{equation}
one needs the appropriate boundary conditions to solve this
equation for $U(x,y)$. To solve the equation on a digital
computer, it is usual to discretize the equation and to work with
a finite lattice. The finite difference version to the Laplace's
equation is obtained by using the well known approximations:
\begin{equation}
 \frac{\partial ^{2}U}{\partial
 x^{2}}=U_{i,j-1}-2U_{i,j}+U_{i,j+1},
\end{equation}
\begin{equation}
 \frac{\partial ^{2}U}{\partial
 y^{2}}=U_{i-1,j}-2U_{i,j}+U_{i+1,j}.
\end{equation}

Here we use $i$ for a row subscript, $j$ for a column subscript,
 and $U$ is the pixel value at each node of the mesh.
Adding these two equations and setting the result equal to zero,
yields the condition that the value at any mesh point must be
equal to the average of its neighbors. This is another way of
defining what is known technically as a `harmonic' function.
Expressed in symbols, using four grid neighbors, this gives the
equation:
\begin{equation}
U_{i,j}=\frac{1}{4}(U_{i-1,j}+U_{i+1,j}+U_{i,j-1}+U_{i,j+1}).
\end{equation}

\clearpage

\clearpage
\begin{figure}[htbp]
\figurenum{1} \hspace{-0.5cm}{\plotone{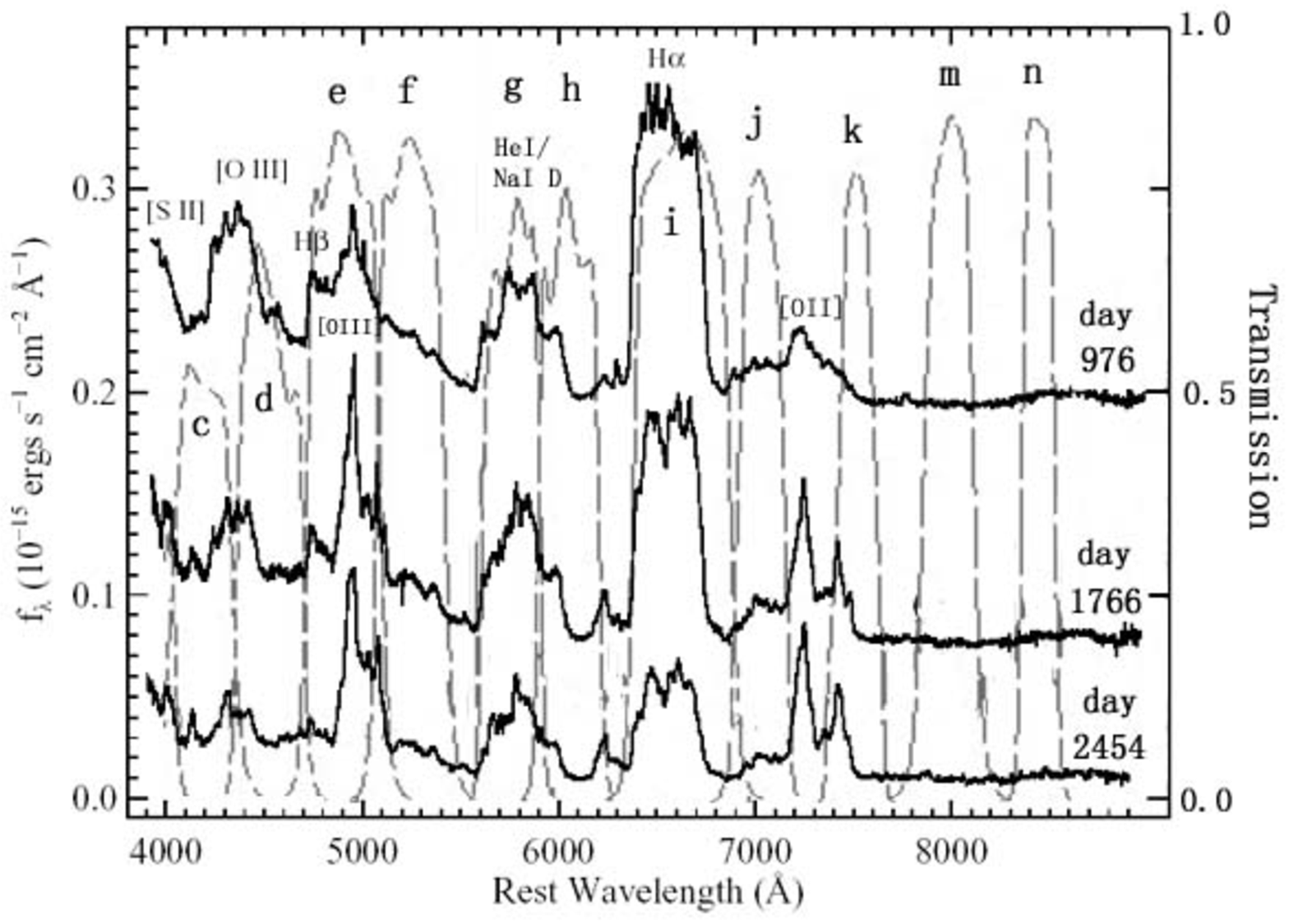}}
\caption{The matching of BATC intermediate-band filters with the
SN 1993J's spectra (the late-time spectra are taken from Matheson
et al. 2000b).} \label{fig:one}
\end{figure}

\clearpage

\begin{figure}
\figurenum{2}
\includegraphics[angle=-90,width=160mm]{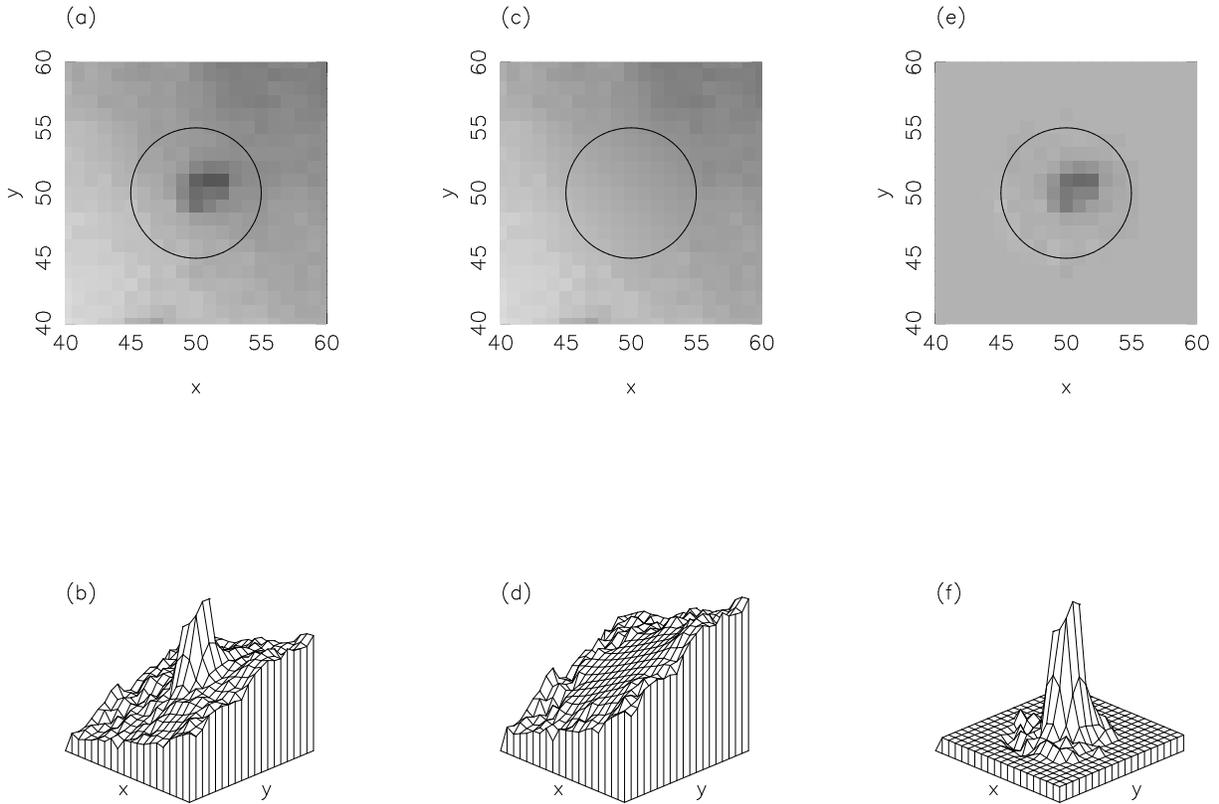}
\caption{Comparison of the i-band image (and mesh), and the
residual image (and mesh) after the fitting background
subtraction. Plotted are $(a)$ The original image of SN1993J;
$(b)$ The mesh of corresponding zone we see in image $(a)$; $(c)$
The fitting background image of the SN; $(d)$ The mesh of
corresponding zone we see in image $(c)$; $(e)$ The image of SN
1993J after subtracting the fitting background $(d)$; $(f)$ The
mesh of corresponding zone we see in image$(e)$. \label{fig:two}}
\end{figure}

\clearpage

\begin{figure}[htbp]
\figurenum{3} \hspace{-0.5cm}{\plotone{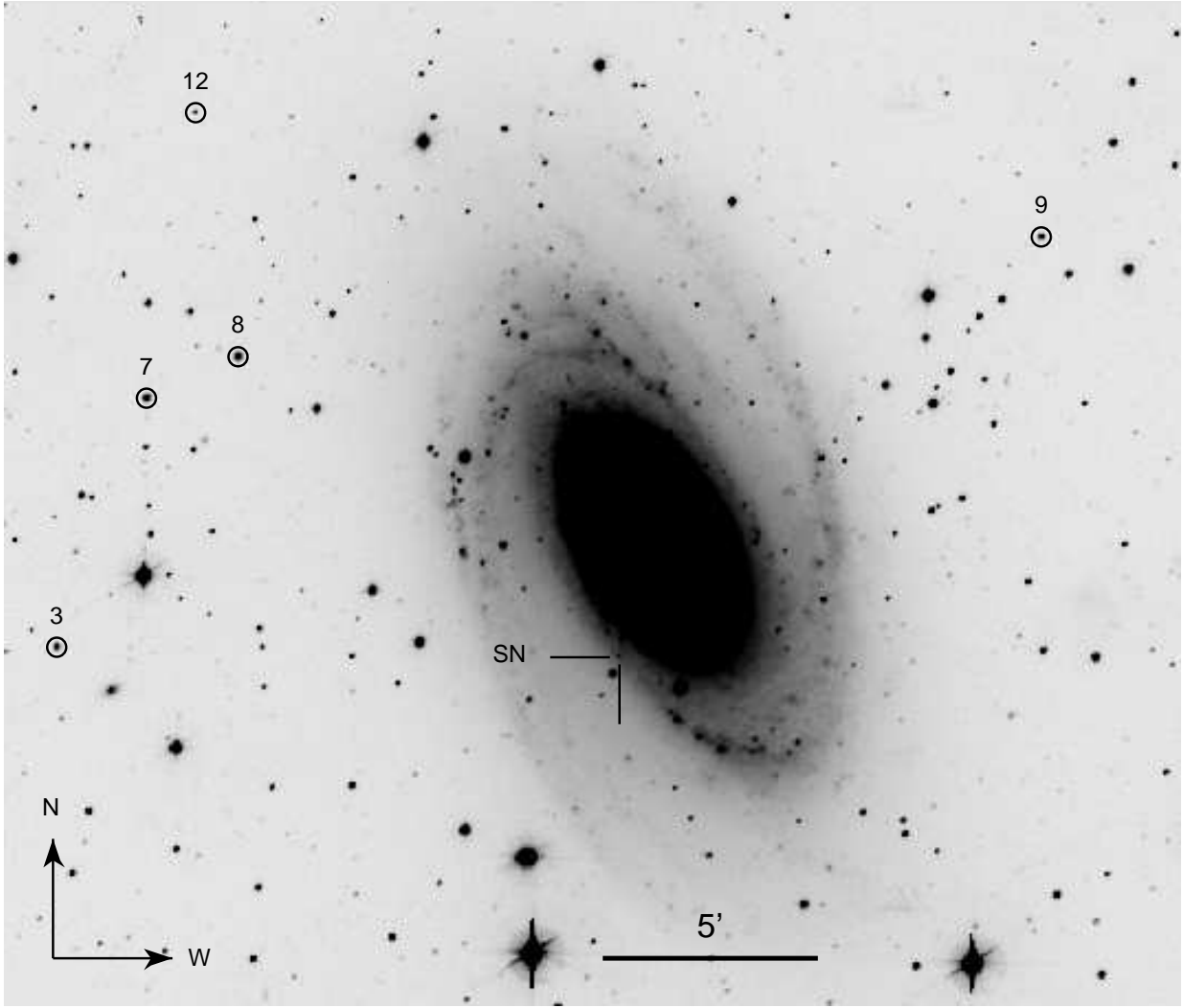}}
\caption{i-band BATC image of the field of SN 1993J (M81), taken
on 1995 Dec 25. The field of view is $30'\times30'$. The five out
of 30 local standard stars are marked.} \label{fig:three}
\end{figure}

\clearpage

\begin{figure}[htbp]
\figurenum{4a} \hspace{-0.5cm}{\plotone{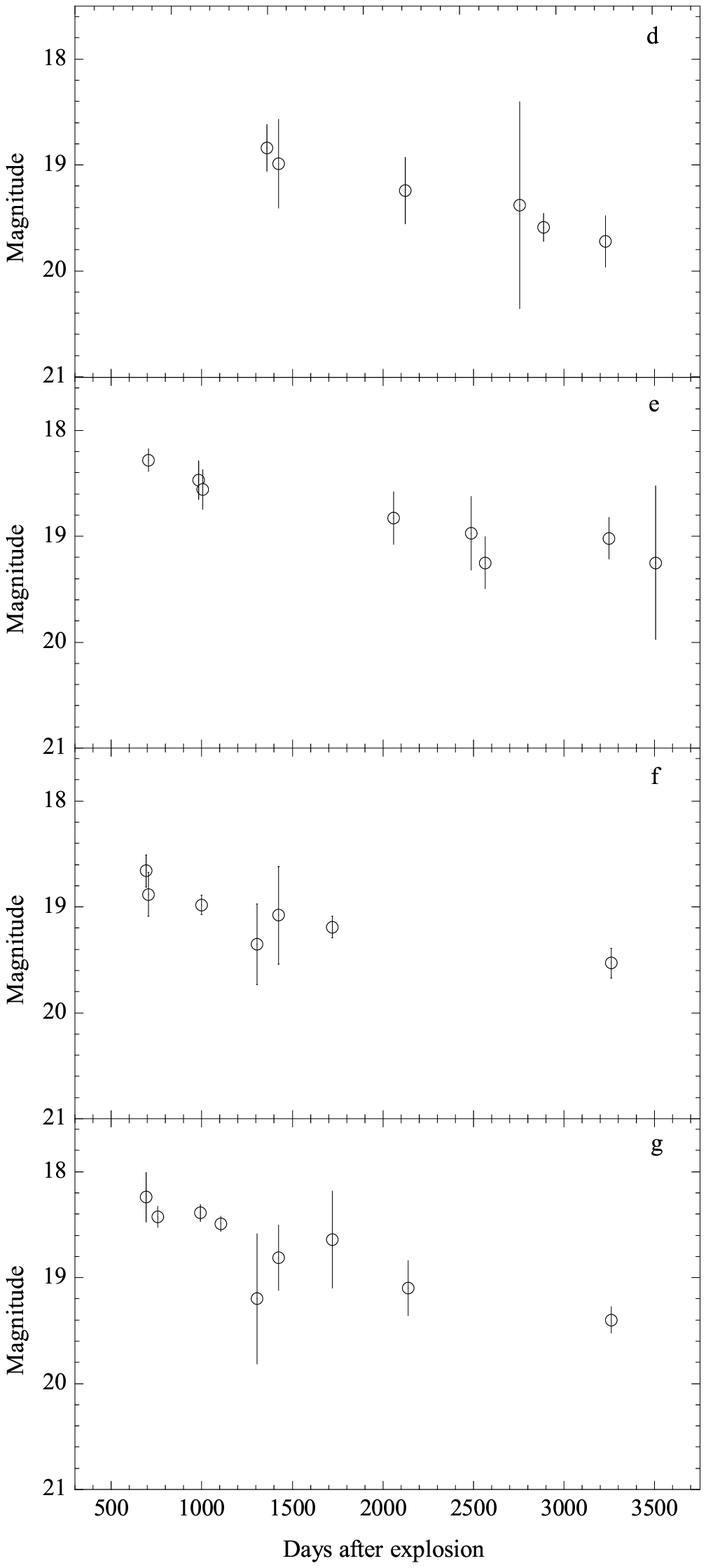}}
\caption{Light curves for SN 1993J in eight intermediate bands.}
\label{fig:foura}
\end{figure}
\clearpage
\begin{figure}[htbp]
\figurenum{4b} \hspace{-0.5cm}{\plotone{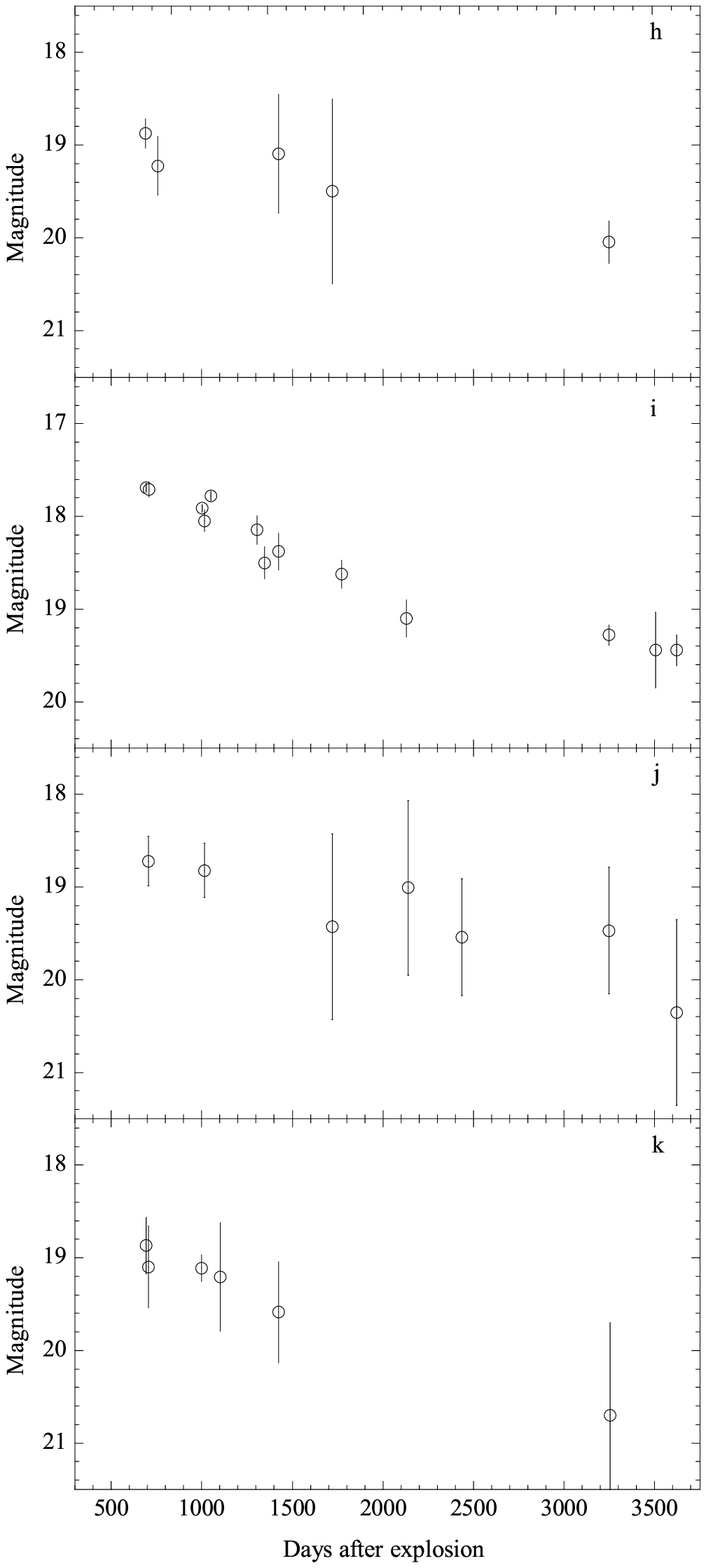}}
\caption{Continued} \label{fig:fourb}
\end{figure}

\clearpage

\begin{figure}[htbp]
\figurenum{5} \hspace{-0.5cm}{\plotone{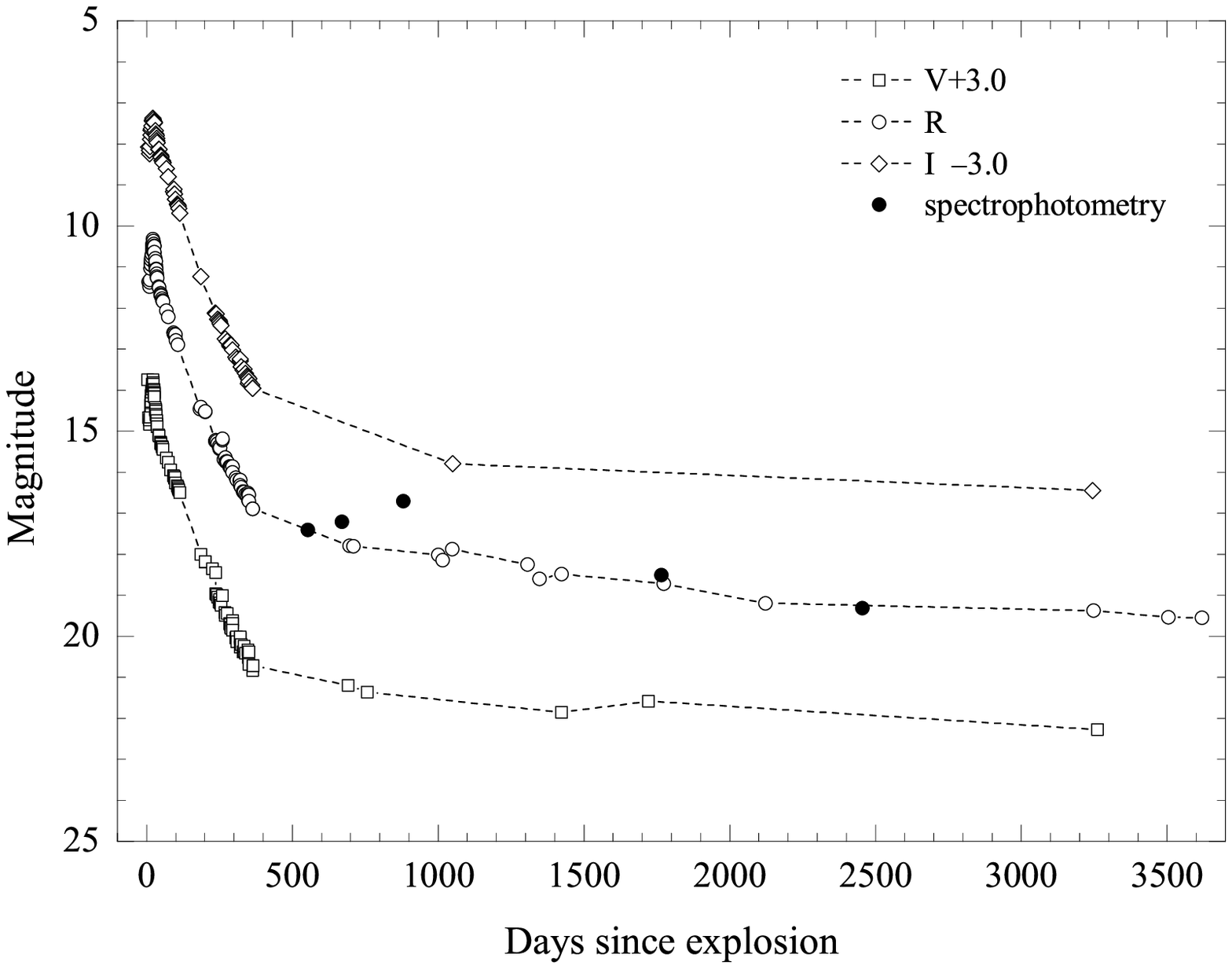}}
\caption{The VRI light curves of SN 1993J over 10 yrs. An offset
has been added to V and I data for clarity. Overplots (filled
circles) are R-band magnitude measured in the late-time spectra
(Matheson et al 2000b).} \label{fig:five}
\end{figure}

\clearpage

\begin{figure}[htbp]
\figurenum{6} \hspace{-0.5cm}{\plotone{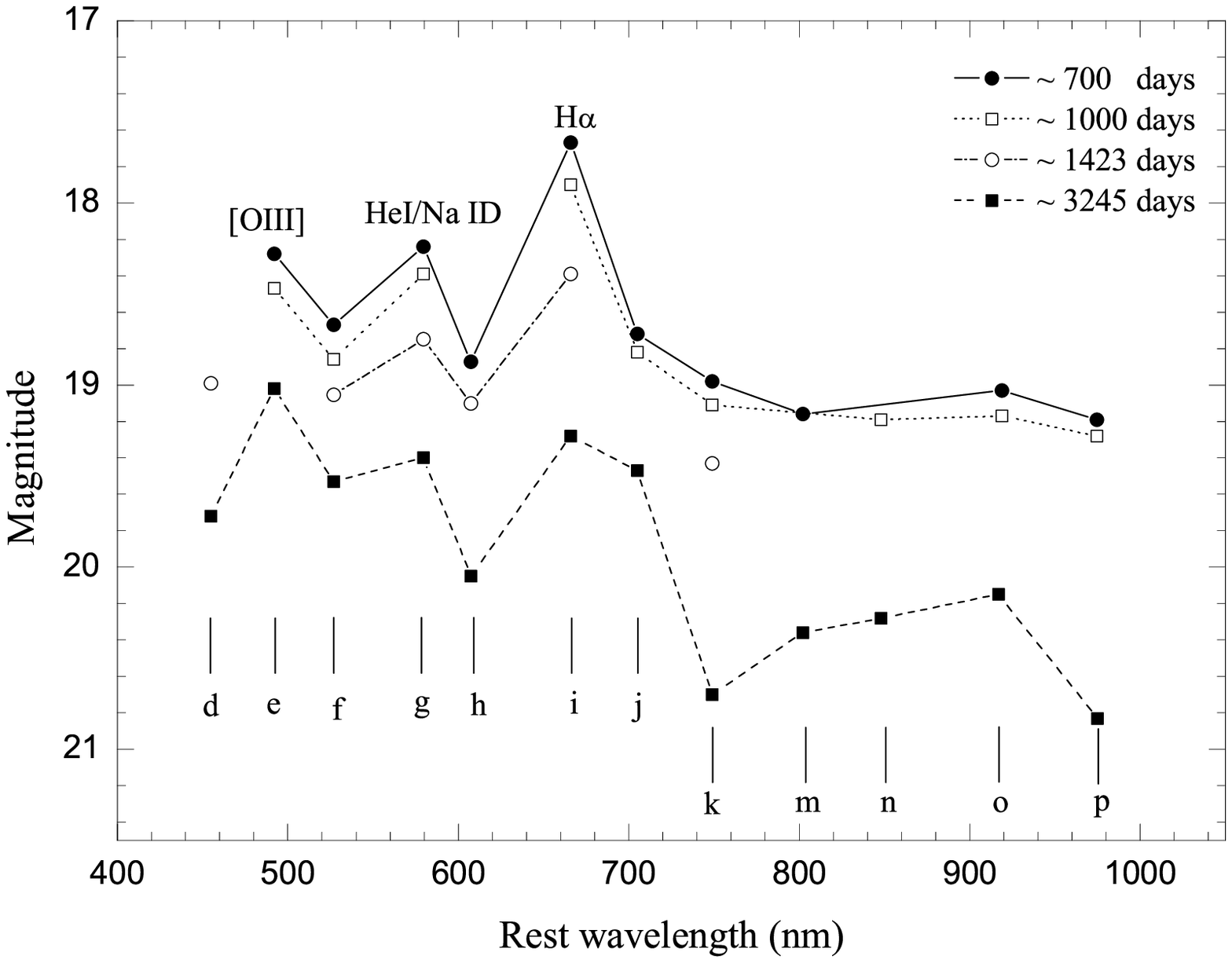}}
\caption{SEDs of SN 1993J at days 700, 1000, 1423 and 3245 after
explosion.} \label{fig:six}
\end{figure}

\clearpage

\begin{figure}[htbp]
\figurenum{7} \hspace{-0.5cm}{\plotone{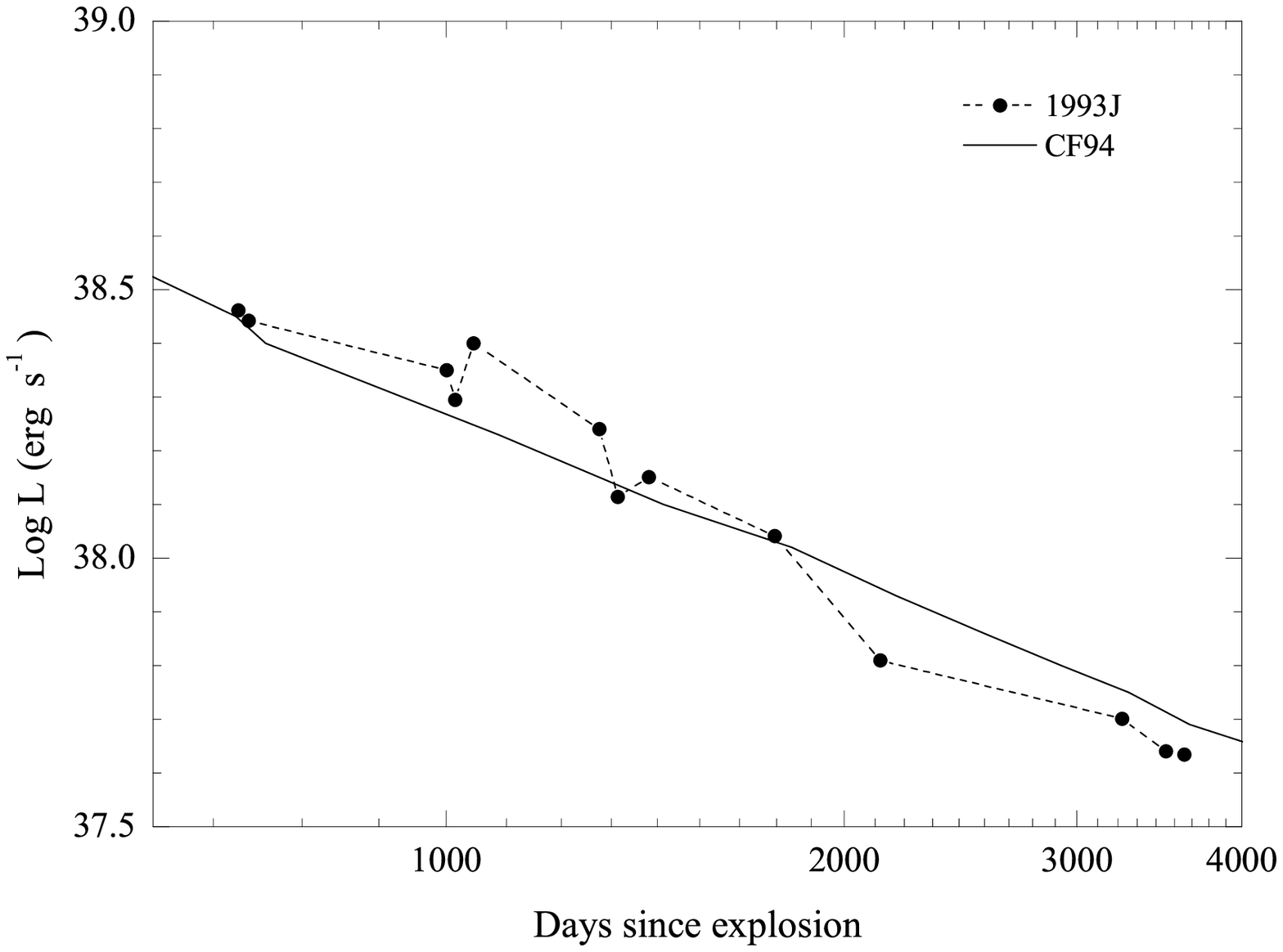}}
\caption{The i-band light curve showing the development of the
$H_{\alpha}$ luminosity. Plotted are (a) the $H_{\alpha}$
luminosity for SN 1993J over the period 1995-2003 (filled
circles), (b) the predicted $H_{\alpha}$ light curve (solid line)
from the power law SN density model of Chevalier $\&$ Fransson
(1994).} \label{fig:seven}
\end{figure}

\clearpage

\begin{deluxetable}{cccccc}
\tablenum{1}
\tablecaption{The details of BATC filters and the observations of
SN 1993J.\label{tbl-1}} \tablewidth{0pt}

\tablehead{ \colhead{Filter name}&
\colhead{$\lambda_{eff}$($\AA$)} & \colhead{FWHM($\AA$)}
&\colhead{Limiting (mag)}&
\colhead{$N_{obs}$\tablenotemark{a}}&\colhead{$N_{phot}$\tablenotemark{b}}
} \startdata
d&4540&332&20.5&6&6\\
e&4925&374&20.0&8&7\\
f&5267&344&20.0&7&4\\
g&5790&289&20.0&9&5\\
h&6074&308&20.0&5&3\\
i&6656&491&20.0&13&8\\
j&7057&238&20.0&7&7\\
k&7546&192&19.0&6&2\\
m&8023&255&19.0&4&2\\
n&8484&167&19.0&3&3\\
o&9182&247&19.0&5&3\\
p&9739&275&18.5&4&4\\
\enddata
\tablenotetext{a}{The number of images.} \tablenotetext{b}{The
number of photometric nights.}
\end{deluxetable}

\clearpage

\tablenum{2}
\begin{sidewaystable}
\caption{Photometry of Comparison Stars \label{tbl-2}}
{\tiny
\begin{tabular}{ccccccccccccccc}
\tableline\tableline ID. & $\alpha$(2000.0)& $\delta$(2000.0) &d
&e &f &g &h &i &j &k &m &n
&o &p\\
\tableline
1&09:51:20.65&68:58:09.3&13.998(04)&13.995(05)&13.779(09)&13.511(09)&13.263(08)&13.338(02)&13.267(08)&13.128(08)&13.169(05)&13.160(08)&13.013(10)&13.025(21)\\
2&09:52:10.73&69:00:10.1&15.366(10)&15.287(10)&15.065(26)&14.808(23)&14.552(23)&14.561(05)&14.476(21)&14.316(23)&14.330(12)&14.222(20)&14.209(28)&14.153(26)\\
3&09:52:55.42&69:01:01.6&14.874(07)&14.849(08)&14.646(17)&14.560(18)&14.329(19)&14.397(05)&14.328(17)&14.152(22)&14.239(10)&14.165(19)&14.116(22)&13.996(17)\\
4&09:59:31.29&69:01:57.3&14.038(04)&14.022(05)&13.811(09)&13.583(10)&13.364(08)&13.422(02)&13.424(10)&13.195(09)&13.330(06)&13.244(09)&13.166(11)&13.141(24)\\
5&09:51:37.55&69:03:52.7&15.669(13)&15.557(12)&15.364(31)&15.038(27)&14.772(26)&14.781(06)&14.797(24)&14.516(28)&14.513(13)&14.476(24)&14.321(29)&14.357(28)\\
6&10:00:30.51&69:07:47.0&16.313(20)&16.301(21)&16.030(56)&15.584(42)&15.232(24)&15.198(08)&15.043(34)&14.848(41)&14.890(19)&14.803(38)&14.714(40)&14.543(30)\\
7&09:53:16.90&69:07:02.6&15.102(08)&14.965(08)&14.831(21)&14.244(16)&13.942(13)&13.892(03)&13.770(12)&13.532(12)&13.543(06)&13.447(10)&13.354(13)&13.339(27)\\
8&09:53:41.00&69:08:06.6&14.193(05)&14.086(05)&13.895(09)&13.685(10)&13.431(09)&13.470(03)&13.421(09)&13.267(09)&13.296(05)&13.214(09)&13.177(11)&13.084(23)\\
9&09:56:44.92&69:09:00.9&15.521(11)&15.505(12)&15.267(28)&15.010(28)&14.759(26)&14.739(06)&14.706(23)&14.514(27)&14.571(14)&14.484(28)&14.406(30)&14.333(33)\\
10&09:57:15.44&69:11:29.4&16.928(34)&16.809(34)&16.605(47)&15.928(34)&15.698(31)&15.681(12)&15.564(27)&15.331(25)&15.341(27)&15.212(28)&14.948(18)&14.971(16)\\
11&09:59:20.78&69:13:19.8&15.619(12)&15.652(13)&15.449(37)&15.240(34)&15.093(37)&15.183(08)&15.102(35)&14.932(23)&15.025(22)&14.918(37)&14.797(45)&14.715(33)\\
12&09:59:03.06&69:13:53.5&16.890(33)&16.858(36)&16.651(32)&16.460(28)&15.951(27)&16.074(26)&16.076(21)&15.660(18)&15.730(28)&15.641(24)&15.519(22)&15.834(30)\\
13&09:53:27.32&69:13:53.6&14.982(08)&14.846(07)&14.712(19)&14.320(15)&13.970(14)&14.005(04)&13.947(12)&13.768(15)&13.826(08)&13.744(15)&13.637(17)&13.572(29)\\
14&09:55:14.89&69:19:26.4&14.223(05)&14.241(05)&14.102(11)&13.974(12)&13.779(13)&13.836(03)&13.814(11)&13.651(14)&13.742(07)&13.709(14)&13.641(17)&13.682(33)\\
15&09:58:12.20&69:20:11.3&14.744(07)&14.750(07)&14.624(17)&14.358(16)&14.158(15)&14.279(04)&14.309(17)&14.076(18)&14.154(10)&14.108(19)&14.050(23)&14.007(25)\\
16&09:58:35.66&69:20:47.4&14.920(08)&14.899(08)&14.789(20)&14.610(19)&14.376(20)&14.456(05)&14.413(19)&14.217(22)&14.354(11)&14.332(23)&14.157(26)&14.298(27)\\
17&09:56:16.90&69:20:43.4&15.441(11)&15.362(11)&15.132(25)&14.954(28)&14.747(28)&14.753(06)&14.700(24)&14.564(29)&14.542(13)&14.540(26)&14.457(31)&14.284(24)\\
18&09:52:54.07&69:21:15.9&14.846(07)&14.783(07)&14.660(19)&14.523(17)&14.311(18)&14.400(04)&14.354(17)&14.189(22)&14.321(10)&14.271(21)&14.134(25)&14.280(28)\\
19&09:56:18.84&69:23:43.7&15.508(12)&15.470(12)&15.344(31)&15.135(27)&14.849(29)&14.937(07)&14.818(25)&14.705(38)&14.769(17)&14.773(29)&14.783(34)&14.520(32)\\
20&09:52:21.53&69:29:30.6&15.478(12)&15.389(11)&15.317(30)&14.989(24)&14.794(28)&14.880(06)&14.783(26)&14.701(35)&14.723(16)&14.616(29)&14.628(38)&14.616(39)\\
21&09:52:09.61&69:29:56.3&17.335(46)&17.178(43)&17.083(35)&16.537(39)&16.625(37)&16.796(37)&16.493(31)&16.497(33)&16.593(34)&16.530(34)&16.219(29)&16.798(45)\\
22&09:58:47.33&69:33:14.1&15.606(12)&15.640(12)&15.499(35)&15.342(35)&15.039(34)&15.160(08)&15.099(33)&14.984(39)&14.940(18)&14.942(42)&14.957(50)&14.767(30)\\
23&09:53:49.38&69:33:56.8&15.591(13)&15.617(12)&15.366(33)&15.405(35)&15.122(39)&15.291(09)&15.236(35)&15.124(35)&15.192(26)&15.133(43)&15.023(25)&15.333(34)\\
24&09:59:56.37&69:39:19.1&14.634(06)&14.690(07)&14.495(17)&14.349(15)&14.161(17)&14.265(04)&14.246(17)&14.015(19)&14.139(10)&14.117(20)&14.072(22)&14.028(48)\\
25&09:58:21.48&69:41:17.9&15.042(08)&14.989(08)&14.789(20)&14.584(20)&14.407(21)&14.477(05)&14.492(19)&14.234(22)&14.297(11)&14.367(24)&14.218(25)&14.125(50)\\
26&10:00:03.31&69:42:44.3&16.437(24)&16.227(19)&15.894(50)&15.214(32)&14.981(31)&14.761(06)&14.632(22)&14.105(20)&14.054(09)&13.915(17)&13.796(18)&13.675(36)\\
27&09:56:49.93&69:43:51.6&17.125(41)&17.137(41)&16.736(35)&16.514(34)&16.164(37)&16.093(16)&15.933(23)&15.935(26)&15.710(34)&15.523(22)&15.456(22)&15.700(35)\\
28&09:50:16.86&69:42:54.6&16.029(17)&16.014(18)&15.900(31)&15.640(20)&15.422(25)&15.509(09)&15.486(29)&15.283(32)&15.322(28)&15.199(20)&15.269(19)&14.967(23)\\
29&09:59:21.70&69:48:37.8&15.739(13)&15.699(13)&15.504(37)&15.253(35)&15.033(35)&15.056(07)&15.017(32)&14.843(37)&14.864(17)&14.873(38)&14.752(39)&14.745(35)\\
30&09:54:34.19&69:48:36.3&15.108(09)&15.158(09)&14.927(22)&14.771(23)&14.571(24)&14.637(05)&14.604(23)&14.438(26)&14.464(13)&14.437(24)&14.438(32)&14.470(27)\\
\tableline
\end{tabular}
} {\small Note. $-$ All quantities are magnitudes. Uncertainties
in the last two digits are indicated in parentheses.}
\end{sidewaystable}

\clearpage

\tablenum{3}
\begin{sidewaystable}
\caption{Photometry of Comparison Stars \label{tbl-2}} {\tiny
\begin{tabular}{lccccccccccccccc}
\tableline\tableline
Date$^{a}$&d&e&f&g&h&i&j\\
\tableline
692 &\nodata &\nodata &18.66(15)&18.24(23)&18.87(16)&17.69(05)&\nodata&18.87(30)&\nodata &\nodata &\nodata&\nodata&18.25&17.79&\nodata\\
708 &\nodata&18.28(11)&\nodata&\nodata&\nodata&17.71(08)&18.72(26)&19.10(44)&19.16(46)&\nodata&19.03(54)&19.19(68)&\nodata&17.81&\nodata\\
757 &\nodata&\nodata&18.88(20)&18.42(10)&19.22(32)&\nodata&\nodata&\nodata&\nodata&\nodata&\nodata&\nodata&18.36&\nodata&\nodata\\
982 &\nodata&18.47(18)&\nodata&18.39(08)&\nodata&\nodata&\nodata&\nodata&\nodata&\nodata&\nodata&\nodata&\nodata&\nodata&\nodata\\
1000&\nodata &18.56(19)&18.96(09)&\nodata&\nodata&17.91(04)&\nodata&19.11(15)&\nodata&19.14(37)&\nodata&\nodata&\nodata&18.01&\nodata\\
1016&\nodata&\nodata&\nodata&\nodata&\nodata&18.05(11)&18.82(29)&\nodata&\nodata&\nodata&\nodata&\nodata&\nodata&18.15&\nodata\\
1050&\nodata&\nodata&\nodata&\nodata&\nodata&17.78(05)&\nodata&\nodata&\nodata&19.19(48)&19.17(43)&19.28(52)&\nodata&17.88&18.80\\
1105&\nodata&\nodata&\nodata&18.49(07)&\nodata&\nodata&\nodata&19.21(58)&19.44(37)&\nodata&18.85(60)&18.95(57)&\nodata&\nodata&\nodata\\
1306&\nodata&\nodata&19.35(38)&19.20(61)&\nodata&18.15(15)&\nodata&\nodata&\nodata &\nodata&\nodata&\nodata&\nodata&18.25&\nodata\\
1358&18.84(22)&\nodata&\nodata&\nodata&\nodata&18.50(18)&\nodata&\nodata&\nodata &\nodata&\nodata&\nodata&\nodata&18.60&\nodata\\
1423&18.99(41)&\nodata&19.08(46)&18.81(30)&19.09(64)&18.38(21)&\nodata&19.43(54)&\nodata &\nodata&\nodata&\nodata&18.85&18.48&\nodata\\
1721&\nodata&\nodata&19.19(10)&18.64(46)&19.50(83)&\nodata &19.43(99)&\nodata&\nodata &\nodata &\nodata&\nodata&18.59&\nodata&\nodata\\
1780&\nodata&\nodata&\nodata&\nodata&\nodata&18.62(15)&\nodata&\nodata&\nodata&\nodata&20.14(96)&\nodata&\nodata&18.72&\nodata\\
2059&\nodata&18.83(25)&\nodata&\nodata&\nodata&\nodata&\nodata&\nodata&\nodata &\nodata&\nodata&\nodata&\nodata&\nodata&\nodata\\
2126&19.24(31)&\nodata&\nodata&19.10(26)&\nodata&19.11(20)& 19.01(94)&\nodata&\nodata &\nodata &\nodata&\nodata&\nodata&19.21&\nodata\\
2436&\nodata&\nodata&\nodata&\nodata&\nodata&\nodata&19.54(63)&\nodata&20.11(82) &\nodata &\nodata&\nodata&\nodata&\nodata&\nodata\\
2487&\nodata&18.97(35)&\nodata&\nodata&\nodata&\nodata&\nodata&\nodata&\nodata &\nodata &\nodata&\nodata&\nodata&\nodata&\nodata\\
2563&\nodata&19.25(41)&\nodata&\nodata&\nodata&\nodata&\nodata&\nodata&\nodata &\nodata &\nodata&\nodata&\nodata&\nodata&\nodata\\
2755&19.38(97)&\nodata&\nodata&\nodata&\nodata&\nodata&\nodata&\nodata&\nodata &\nodata &\nodata&\nodata&\nodata&\nodata&\nodata\\
2888&19.59(13)&\nodata&\nodata&\nodata&\nodata&\nodata&\nodata&\nodata&\nodata &\nodata &\nodata&\nodata&\nodata&\nodata&\nodata\\
3245&19.72(24)&19.02(19)&19.53(14)&19.40(12)&20.05(23)&19.28(11)&19.47(68)&20.70(99)&20.36(74)&20.28(95)&20.15(110)&20.83(126)&19.28&19.38&19.46\\
3504&\nodata&19.15(24)&\nodata&\nodata&\nodata&19.44(41)&\nodata&\nodata&19.44(41)&\nodata &\nodata&\nodata&\nodata&19.54&\nodata\\
3620&\nodata&\nodata&\nodata&\nodata&\nodata&19.44(17)&20.35(100)&\nodata&\nodata&\nodata&\nodata&\nodata&\nodata&19.54&\nodata\\
\tableline
\end{tabular}
} {\small $^{a}$Days after explosion, 1993 March UT27.5.}
\end{sidewaystable}

\clearpage

\begin{deluxetable}{ccc}
\tablenum{4}

\tablecaption{Late-time luminosity decline rates of SN 1993J in 12
intermediate passbands. \label{tbl-4}} \tablewidth{0pt}
\tablehead{
\colhead{Band} &\colhead{Phase\tablenotemark{a} (days)}& \colhead{Decline rate (mag 100d$^{-1}$)}\\
} \startdata
d &1358-3245&$0.046\pm0.014$\\
e &708-3504&$0.033\pm0.007$\\
f &692-3245&$0.030\pm0.006$\\
g &692-3245&$0.043\pm0.006$\\
h &692-3245&$0.044\pm0.011$\\
i &692-3600&$0.066\pm0.004$\\
j &708-3600&$0.041\pm0.021$\\
k &692-3245&$0.071\pm0.037$\\
m &708-3245&$0.048\pm0.031$\\
n &1000-3245&$0.050\pm0.045$\\
o &708-3245&$0.054\pm0.046$\\
p &708-3245&$0.067\pm0.056$\\
\enddata
\tablenotetext{a}{Days after explosion, 1993 March UT27.5.}
\end{deluxetable}

\begin{deluxetable}{lcccccc}
\tablenum{5A}
\tablecaption{Observed line flux from SN 1993J.} \tablewidth{0pt}
\tablehead{\multicolumn{5}{r}{Age (year)}\\
\cline{2-7}

\colhead{Emission Line}&\colhead{1.9}&\colhead{2.7}&
\colhead{4.8}&\colhead{5.7}&\colhead{8.9}&\colhead{9.6} }

\startdata
$H_{\alpha}$\tablenotemark{a}  &2.84 &2.10 &1.03    &0.63 &0.45 &0.41\\
Na I D+He I5876                &0.60 &0.71 &0.86    &1.12 &1.06 &\nodata\\
$[O III]$4959-5007+$H_{\beta}$ &0.58 &0.69 &\nodata &1.50 &1.69 &1.72\\
\enddata
\end{deluxetable}

\begin{deluxetable}{lcccc}
\tablenum{5B}

\tablecaption{Observed line flux from SN 1993J.} \tablewidth{0pt}
\tablehead{\multicolumn{4}{r}{Age (year)}\\
\cline{2-5} \colhead{Emission Line}&\colhead{2}&\colhead{5}&
\colhead{10}&\colhead{10(RSG)\tablenotemark{b}} }

\startdata
$H_{\alpha}$\tablenotemark{a}  &0.96 &0.19 &0.09    &0.49 \\
Na I D                         &0.38 &0.61 &1.00    &0.17 \\
$[O III]$4959-5007             &0.27 &1.00 &2.20 &3.40 \\
\enddata

\tablenotetext{a}{The luminosity of $H_{\alpha}$ in $10^{38}$ erg
$s^{-1}$.} \tablenotetext{b}{RSG is the red supergiant model at 10
yr.}
\end{deluxetable}

\end{document}